\documentclass[12pt,preprint]{aastex}
\newtheorem{theorem}{Theorem}
\newtheorem{lemma}{Lemma}
\newtheorem{corollary}{Corollary}
\newenvironment{proof}{~ \\[0.1in] {\bf Proof.} }{\hfill $\Box$ \bigskip \\[0.1in]}

\usepackage[intlimits]{amsmath}
\usepackage{times}

\newcommand{\xt}{\tilde{x}}
\newcommand{\xtt}{\tilde{\tilde{x}}}
\newcommand{\dxt}{\Delta\tilde{x}}
\newcommand{\thetat}{\tilde{\theta}}
\newcommand{\sroot}{ \sqrt{(\xt-x)^2 + (g(\xt)-h(x))^2} }

\newcommand{\Psf}{\text{Psf}}

\newcommand{\thetab}{\boldsymbol{\theta}}
\newcommand{\thetatb}{\boldsymbol{\tilde{\theta}}}
\newcommand{\xib}{\boldsymbol{\xi}}
\newcommand{\xb}{\boldsymbol{\tilde{x}}}
\newcommand{\ds}{\displaystyle}
\newcommand{\half}{{\text{\scriptsize$\frac{1}{2}$}}}
\newcommand{\halfd}{{\text{\scriptsize$\frac{d}{2}$}}}

\shorttitle{Two-Mirror Apodization}
\shortauthors{Traub and Vanderbei}

\begin{document}


\title{Two-Mirror Apodization for High-Contrast Imaging}


\author{Wesley A. Traub}
\affil{Harvard-Smithsonian Center for Astrophysics}
\email{wtraub@cfa.harvard.edu}

\and

\author{Robert J. Vanderbei}
\affil{Operations Research and Financial Engineering, Princeton University}
\email{rvdb@princeton.edu}

\begin{abstract}
Direct detection of extrasolar planets will require imaging systems capable
of unprecedented contrast.  Apodized pupils provide an attractive way to
achieve such contrast but they are difficult, perhaps impossible, to
manufacture to the required tolerance and they absorb about $90\%$ of the
light in order to create the apodization, which of course lengthens the exposure
times needed for planet detection.  A recently proposed alternative is to
use two mirrors to accomplish the apodization.  With such a system, 
no light is lost.  In this paper, we provide a careful mathematical
analysis, using one dimensional mirrors,
of the on-axis and off-axis performance of such a two-mirror 
apodization system.  There appear to be advantages and disadvantages
to this approach.  In addition to not losing any light,
we show that the nonuniformity of the apodization implies an extra
magnification of off-axis sources and thereby makes it possible to build a
real system with about half the aperture that one would otherwise require
or, equivalently,
resolve planets at about half the angular separation as one can achieve with
standard apodization.  More specifically, ignoring pointing error and stellar
disk size, a planet at $1.7 \lambda/D$ ought to be at the edge of detectability.
However, we show that the non-zero size of a stellar disk 
pushes the threshold for high-contrast so that 
a planet must be at least $2.5 \lambda/D$ from its
star to be detectable.
The off-axis analysis of two-dimensional mirrors is left for future study.
\end{abstract}


\section{Introduction}
\label{sec:intro}

In a recent paper, \citet{GOPP02} propose using two mirrors to convert
a uniform-intensity beam to one of nonuniform intensity---that is, an
apodized exit beam.  The motivation for their work was to ``mask''
the secondary mirror in designs
of large two-mirror telescopes with fast spherical primaries.
As an application of the idea, they point to radio and submillimeter-wave
telescopes since for these it is important to suppress sidelobes and reduce
background noise.
However, its applicability to high-contrast imaging in the context
of the Terrestrial Planet Finder (TPF) project immediately caught our attention 
and the result is the work described in this paper.

Independently of Goncharov et. al., 
\citet{Guy03} studied the same two-mirror
apodization idea and analyzed its applicability to extra-solar planet finding.
His work was motivated by Labeyrie's paper on pupil densification
(\cite{Lab96}).  The idea behind pupil densification is to reduce the
intensity of the diffraction side lobes 
%by remapping the sparse pupil of an interferometer.
by remapping the sparse input pupil of an interferometer to a more
compact exit pupil.  Pupil densification provides the advantage of a
better-concentrated image at the cost of field of view, since it
violates the ''golden rule" of combining beams (\cite{Tra86});
this rule states that, to provide a large field of
view, the exit pupil of an optical system must be a scaled version of the
input pupil.  Since pupil densification is a special case of pupil
remapping, both will suffer field of view loss if the scaling factor is
not constant.

   Nevertheless, the potential gains from pupil mapping are so
significant that it is worthwhile exploring this technique for the purpose
of extrasolar planet searching, and this is the goal of the present paper.

In his paper, Guyon argues
that the two-mirror approach provides a system with (i) a smaller inner
working angle ($1 \lambda/D$ vs. $4 \lambda/D$) and consequently 
a smaller overall aperture and (ii) a higher throughput with a corresponding
reduction in exposure time.  
Here, and throughout the paper, $\lambda$ is wavelength and $D$ is the
diameter of the primary mirror.
Guyon also addresses, at least qualitatively,
the issue of off-axis performance of the system.  

In this paper, we
give a careful mathematical analysis of the on-axis and off-axis performance
of this two-mirror system.  We deal exclusively with one-dimensional mirrors
in two-dimensional space.  While the on-axis analysis is easily extended to 
real two-dimensional mirrors in three space dimensions, the off-axis analysis,
which is the crux of this paper, is much more difficult in this higher
dimension.  Hence, we leave such analysis to future work.

The first and most obvious advantage of the two-mirror system is that 
throughput is increased since no light is lost.  Furthermore,
an interesting and unexpected advantage of this system is 
that there is an inherent magnification introduced by 
a nonuniform apodization.  For example, a planet at $5 \lambda/D$ in the 
sky actually appears at about $12 \lambda/D$ in the image plane.  
This improves the effective inner working angle by a factor of $2.4$ and
explains, we believe, Guyon's claims about improved inner working angle.
Consequently, such a system for planet finding could be implemented with a
smaller mirror than would otherwise be required or, equivalently, could detect
planets less than half as far from their parent star with the same aperture.

There is, however, a downside.  We show that the inner working
angle of an ideal system (with an on-axis source) is still $4 \lambda/D$
and furthermore that even tiny off-axis light sources, such as arising from
the nonzero size of the stellar disk, increase the inner working angle
to at least $6 \lambda/D$.  So, a star at $2.5 \lambda/D$ in the
sky will appear at $6 \lambda/D$ in the image plane and hence lie on the edge
of detectability.  This still represents a significant improvement over the
$4 \lambda/D$, which is the best one can do with standard apodization or with
pupil masks.

In the next section, we derive the basic equations relating the shapes
of two one-dimensional mirrors to the apodization they produce.  Then,
in Section \ref{sec:oap}, we give a careful analysis of the off-axis
performance of this two-mirror system.  Section \ref{sec:simpex}
then goes through
some specific examples, including the case of two parabolas, where everything
can be computed explicitly.  Finally, in Section \ref{sec:numex}, we show the
results of numerical computations using the known optimal apodizer.

We end the present section with a word about our notation.  
Given any function of a
single real variable, we use a prime to denote the derivative of this function
with respect to its variable.  So, for example, $h'(x) = dh(x)/dx$.

\section{Mirror-Based Apodization}
\label{sec:mba}

\begin{figure}
\begin{center}
%\text{\includegraphics[width=4.0in]{C:/user/rvdb/public_html/JAVA/tpf/apodmirror/plot.pdf}}
\text{\includegraphics[width=4.0in]{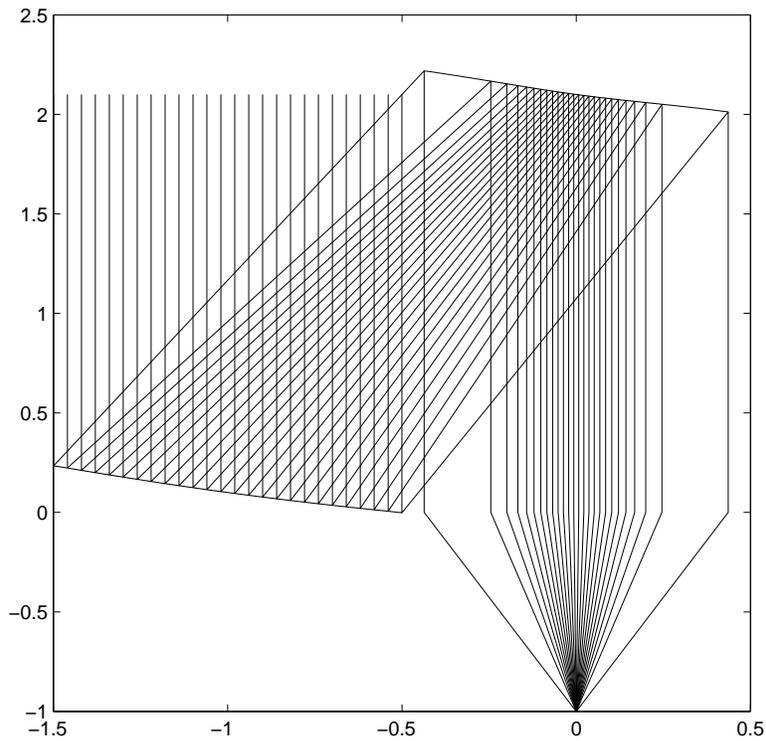}}
\end{center}
\caption{Parallel light rays come down from above, reflect off the bottom
mirror (on the left), bounce up to the top mirror (on the right), 
and then exit downward
as a parallel bundle with a concentration of rays in the center
of the bundle and thinning out toward the edges---that is, the
exit bundle is apodized.
A focusing element on the $x$-axis then creates an image.
%(Note that, as drawn, the focusing element interferes with
%the optical path between the two mirrors.  Of course, in practice
%the focusing element could be shifted down out of the way.)
}
\label{fig:twomirrors}
\end{figure}

Figure \ref{fig:twomirrors} illustrates the use of two mirrors to
produce an apodization.
Parallel light rays come down from above, reflect off the bottom
mirror (on the left), bounce up to the top mirror (on the right), 
and then exit downward
as a parallel bundle with a concentration of rays in the center
of the bundle and thinning out toward the edges---that is, the
exit bundle is apodized.
A focusing element on the $x$-axis then creates an image.

Let $h(x)$, $a \le x \le b$,
denote the height-function defining the lower mirror and
let $g(\xt)$, $-\halfd \le \xt \le \halfd$,
denote the height-function defining the upper mirror.
We suppose there is a one-to-one correspondence between $x$'s 
in the interval $[a,b]$ and the interval $[-\halfd,\halfd]$
that is determined by the desired apodization.
Let $\xt(x)$ denote the value in $[-\halfd,\halfd]$ associated with $x$ in 
$[a,b]$.
We assume that the function $\xt$, which we call the {\em transfer function},
is strictly monotone (either increasing or decreasing).  
It then has a functional inverse.  
Let $x(\xt)$ denote this inverse function.
The {\em apodization} $A(\cdot)$ associated with this correspondence
is given by
\begin{equation} \label{11}
    \frac{dx}{d\xt}(\xt) = A(\xt) .
\end{equation}
Adding a boundary condition, such as $x(-\halfd) = a$, makes it possible to
derive the correspondence from the apodization.

The shapes of the two mirrors are determined by enforcing 
equality between angle-of-incidence and angle-of-reflection
(i.e., Snell's law) at the two mirror surfaces.
The tangent vector to the first mirror is $(1, h'(x))$.  The
unit incidence vector is $(0,1)$ and the unit reflection vector is
$(\xt-x, g(\xt)-h(x))/S(x,\xt)$, where
\begin{equation}
    S(x,\xt) = \sroot 
\end{equation}
denotes the distance between the point $(x,h(x))$ on the first mirror and
$(\xt, g(\xt))$ on the second mirror.
The scalar product between the tangent vector and the unit incidence vector
must be equal to the negative of the scalar product between the tangent
vector and the unit reflection vector.
Equating these inner products, we get the following
differential equation for $h$:
\begin{equation} \label{10}
    h'(x) = \frac {x-\xt} {S(x,\xt) - h(x) + g(\xt)} .
\end{equation}
Applying the same technique to the surface of the second mirror at $\xt$,
we get the same expression for $g'$ and so
\begin{equation} \label{12}
    g'(\xt) = h'(x) .
\end{equation}
Equation \eqref{12} says that, for a given ray, the slopes of the mirrors are
locally the same, which  is to be expected since the input ray is parallel to
the output ray.
From equations \eqref{10} and \eqref{12}, 
it appears that the differential equations for $h$ and $g$ are a coupled
system of ordinary differential equations.  Our first task is to show how to
decouple them.

Let 
\begin{equation} \label{14}
    P_0(x,\xt) = S(x,\xt) + g(\xt) - h(x) ,
\end{equation}
denote the denominator in \eqref{10}.
It is the total {\em optical path length} from a point on any given horizontal
line drawn above the first mirror (and below the second)
vertically down to the first mirror diagonally up to the second mirror
and finally vertically back down to the same horizontal line
below the second mirror.

\begin{theorem} \label{thm1}
    The optical path length $P_0$ is a conserved quantity for on-axis rays.
\end{theorem}
\begin{proof}
    If we regard $x$ as a function of $\xt$ and differentiate with respect to
    $\xt$, we get
    \begin{equation}
        \frac{d}{d\xt} \left( S(x,\xt) + g(\xt) - h(x) \right)
        =
        G(\xt) - H(\xt) \frac{dx}{d\xt} ,
    \end{equation}
    where
    \begin{eqnarray} 
        G(\xt) & = &
        \frac{
            \xt - x + (g(\xt)-h(x))g'(\xt)
        }{
            S(x,\xt)
        }
        + g'(\xt) \label{15}
        \\[0.2in]
        H(\xt) & = &
        \frac{
            \xt - x + (g(\xt)-h(x))h'(x)
        }{
            S(x,\xt)
        }
        + h'(x) .\label{16}
    \end{eqnarray}
    From \eqref{12}, we see that $G(\xt) = H(\xt)$.
    Using \eqref{10} to eliminate $h'(x)$ from \eqref{16}, we get
    after some straightforward algebraic manipulation that
    $H(\xt) = 0$
%    \begin{eqnarray*}
%        H(\xt) & = & 
%	0
%        \frac{
%            \xt - x - (g(\xt)-h(x))\frac{\xt-x}{S(x,\xt)-h(x)+g(\xt)}
%        }{
%	    S(x,\xt)
%        }
%        - \frac{\xt - x}{S(x,\xt)-h(x)+g(\xt)}
%%        \\
%%        & = &
%%	(\xt - x)
%%        \frac{
%%            S(x,\xt) - h(x) + g(\xt) - (g(\xt) - h(x)) - S(x,\xt)
%%        }{
%%            S(x,\xt) (S(x,\xt) - h(x) + g(\xt))
%%        }
%%        \\
%%        & = &
%%        0 
%	 .
%    \end{eqnarray*}
%    Cross multiplying by the denominators, it is easy to check that $H(x)$
%    vanishes.
    and therefore $P_0$ is a constant.
\end{proof}
Henceforth, we write $P_0$ for this optical-path-length invariant.
\begin{corollary} \label{cor1}
    The differential equations for $h$ and $g$ are decoupled:
    \begin{equation} \label{13}
         h'(x) = g'(\xt) = \frac{x - \xt}{P_0} .
    \end{equation}
\end{corollary}
Isolating $S(x,\xt)$ on one side of the equation in 
\eqref{14} and squaring, we can solve for the difference
between $g(\xt)$ and $h(x)$:
\begin{equation} \label{25}
    g(\xt) - h(x) = -\frac{(\xt-x)^2}{2P_0} + \frac{P_0}{2} .
\end{equation}
Hence, it is only necessary to solve one of the differential equations.
The other mirror surface can then be computed from this difference.

\section{Off-Axis Performance}
\label{sec:oap}

\begin{figure}
\begin{center}
%\text{\includegraphics[width=4.0in]{C:/user/rvdb/public_html/JAVA/tpf/apodmirror/notations.pdf}}
%\text{\includegraphics[height=6in,width=4.0in]{f2.eps}}
\text{\includegraphics[width=6.0in]{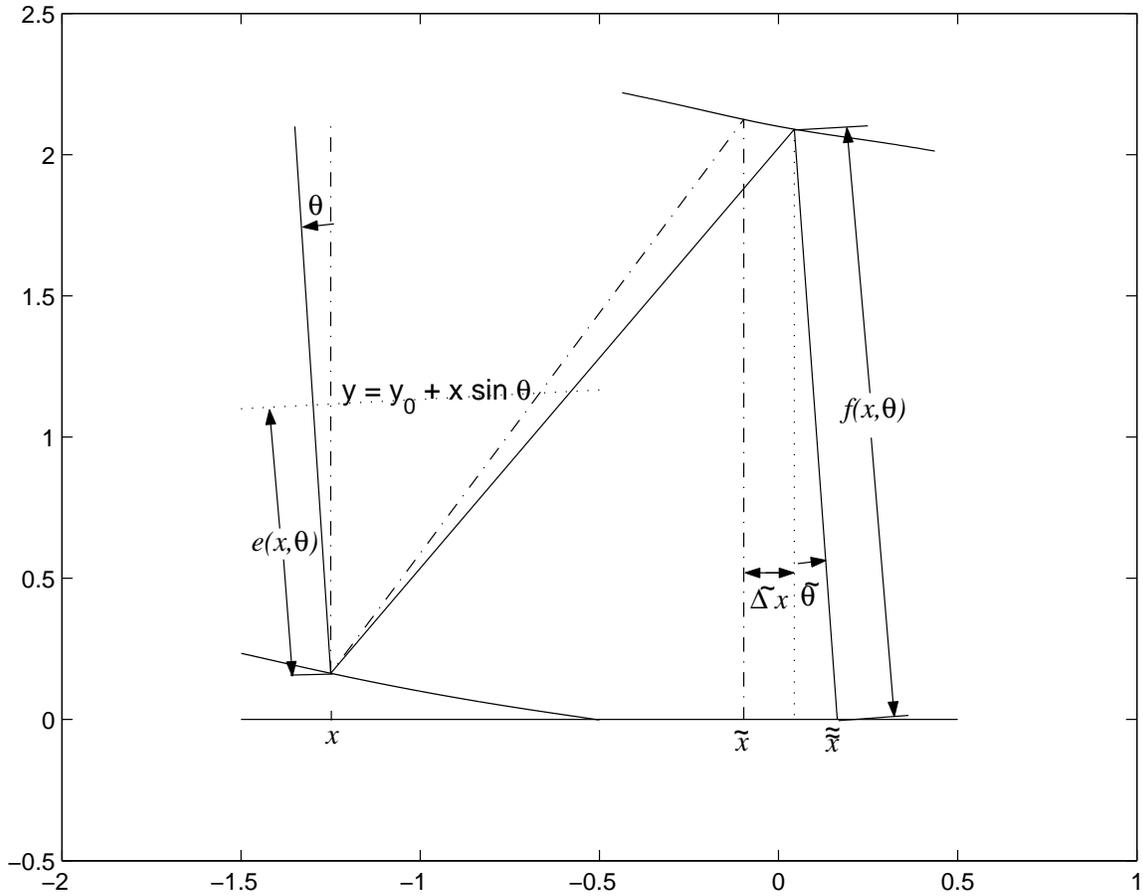}}
\end{center}
\caption{Notations for off-axis expressions.
Parallel light rays come down at an angle $\theta$ from vertical, 
reflect off the bottom
mirror (on the left), bounce up to the top mirror (on the right), 
and then exit downward at an angle $\thetat$ hitting the $x$-axis
at a new location $\xtt$.  
%(Note: it should be regarded as a coincidence
%that the on-axis ray appears to enter the system at $x=-1$.)
}
\label{fig:notations}
\end{figure}

Suppose now that coherent light enters at angle $\theta$ from on-axis
as shown in Figure \ref{fig:notations}.
The light that strikes at position $(x,h(x))$ on the first mirror reflects
off at a different angle and strikes the second mirror at a new position
shifted by $\dxt$ from the on-axis strike point: $(\xt + \dxt, g(\xt+\dxt))$.
Light then reflects off the second mirror at angle $\thetat$ and intersects
the $x$-axis at position $\xtt$.  
We need to compute how $\dxt$, $\thetat$, and $\xtt$ depend on $\theta$ with
the ultimate goal of using these expansions to compute the image-plane
point-spread function as a function of $\theta$.

\begin{lemma} \;\;\; $ \dxt = S(x,\xt) \; \theta + o(\theta) $. \label{thm2}
\end{lemma}
\begin{proof} 
  The unit incidence vector is $(-\sin\theta, \cos\theta)$ and the 
  unit reflection vector is $(\xt+\dxt-x, g(\xt+\dxt)-h(x))/S(x,\xt+\dxt)$.
%  where 
%  \begin{equation}
%    S_{\theta}(x,\xt+\dxt) = \sqrt{(\xt+\dxt-x)^2 + (g(\xt+\dxt)-h(x))^2}.
%  \end{equation}
  As before, Snell's law can be expressed by equating the inner product
  between the unit incidence vector and the mirror's tangent vector with
  the negative of the corresponding inner product for the unit reflection
  vector.  We get that
  \begin{equation} \label{40}
    -\sin\theta + h'(x)\cos\theta
    =
    -\frac{\left(\xt+\dxt-x + h'(x)(g(\xt+\dxt)-h(x))\right)}{S(x,\xt+\dxt)}.
  \end{equation}
  Linearizing the left-hand side in $\theta$ and the right-hand side in
  $\dxt$ and noting that the constant terms in the two linearizations match,
  since they give the on-axis equation satisfied by $h'(x)$, we get
  \begin{equation} \label{41}
      \theta 
      \approx
      \left(
        \frac{1+h'(x)g'(\xt)}{S(x,\xt)}
	-
	\frac{\xt-x+h'(x)(g(\xt)-h(x))}{S(x,\xt)^3}
	\left(\xt-x+g'(\xt)(g(\xt)-h(x))\right)
      \right) \dxt .
  \end{equation}
  Using Corollary \ref{cor1}, the first term on the right simplifies to
  $(1+h'(x)^2)/S(x,\xt)$ and the second term simplifies to 
  $h'(x)^2/S(x,\xt)$.  Hence, we get that to first order 
  $\theta = \dxt/S(x,\xt)$ from which the lemma follows.
\end{proof}

\begin{lemma} \;\;\; \label{thm3}
$ \thetat = \ds \frac{A(\xt)}{S(x,\xt)} \; \dxt + o(\dxt) 
          = A(\xt) \theta + o(\theta).  $
\end{lemma}
\begin{proof} 
  The proof is similar to that of the previous lemma.
  The unit reflection vector is $(\sin\thetat, -\cos\thetat)$ and the 
  unit incidence vector is just the negative of what it was before:
  $-(\xt+\dxt-x, g(\xt+\dxt)-h(x))/S(x,\xt+\dxt)$.
  The mirror's tangent vector at $\xt+\dxt$ is $(1,g'(\xt+\dxt))$.  
  As before, Snell's law can be expressed by equating the inner product
  between the unit incidence vector and the mirror's tangent vector with
  the negative of the corresponding inner product for the unit reflection
  vector.  
  We get that
  \begin{equation} \label{42}
    \sin\thetat - g'(\xt+\dxt)\cos\thetat
    =
    \frac{\left( \xt+\dxt-x + g'(\xt+\dxt) 
                 (g(\xt+\dxt)-h(x)) \right)}{S(x,\xt+\dxt)} .
  \end{equation}
  Linearizing both sides in $\thetat$ and 
  $\dxt$ and noting that the constant terms in the two linearizations match,
  since they give the on-axis equation satisfied by $g'(\xt)$, we now get
  after employing Corollary \ref{cor1} that
  \begin{equation} \label{43}
      \thetat - g''(\xt)\dxt 
      \approx
      \left(
        \frac{1 + g'(\xt)^2 + (g(\xt)-h(x))g''(\xt)}{S(x,\xt)}
	-
        \frac{g'(\xt)^2}{S(x,\xt)}
      \right) \dxt .
  \end{equation}
  Moving $\dxt$ terms to the right-hand side and simplifying, we get
  \begin{eqnarray} 
      \thetat 
      & = &
      \frac{ 1 + g''(\xt) \left( S(x,\xt) + g(\xt) - h(x) \right)}{S(x,\xt)}
      \dxt 
      \nonumber
      \\
      & = &
      \frac{ 1 + g''(\xt) P_0}{S(x,\xt)}
      \dxt 
      \label{44}
  \end{eqnarray}
  Differentiating the expression for $g'(\xt)$ given in Corollary \ref{cor1},
  we get that
  \begin{equation} \label{45}
      g''(\xt) = \frac{dx/d\xt - 1}{P_0} = \frac{A(\xt) - 1}{P_0} .
  \end{equation}
  Substituting \eqref{45} into \eqref{44}, we get the claimed first-order
  relationship between $\thetat$ and $\dxt$.  The relationship between
  $\thetat$ and $\theta$ then follows from Theorem \ref{thm2}.
\end{proof}

From these expansions, we can compute the phase of the wavefront at the
exit pupil or, equivalently, the path length from an isophase line at the
entrance pupil through the system to the exit pupil.  
To this end, note that, for any $y_0$,
the line $y = y_0 + x \sin \theta$, $x \in [b,c]$, is an iso-phase line for the
entering wave.  Fix such a $y_0$.
Let $e(x,\theta)$ denote the length of the light ray
from this line to the first mirror surface.  
The interval $[-\halfd,\halfd]$ of the $x$-axis can be regarded
as an exit pupil.  Let $\xtt$ denote the position along this pupil where
the off-axis light beam exits the system.
Let $f(x,\theta)$ denote the length of the light ray from $(\xt+\dxt, g(\xt+\dxt))$
to this point on the pupil.  The path length from the iso-phase line to the
exit pupil is 
\begin{equation} \label{55}
    P(x) = e(x,\theta) + S(x,\xt+\dxt) + f(x,\theta) .
\end{equation}

\begin{theorem} \;\;\; \label{thm4}
  $
    P(x) = P_0 + y_0 + x \theta + o(\theta)  .
  $
\end{theorem}
\begin{proof}
  Simple geometry reveals that
  \begin{equation} \label{207}
      e(x,\theta) = y_0 -h(x) \cos \theta + x \sin \theta 
           = y_0 -h(x) + x \theta + o(\theta)
  \end{equation}
  and that
  \begin{eqnarray}
      f(x,\theta) 
      & = & \frac{g(\xt+\dxt)}{\cos \thetat} 
        =   g(\xt) + g'(\xt) \dxt + o(\dxt) \nonumber \\
      & = & g(\xt) + g'(\xt) S(x,\xt) \theta + o(\theta) . \label{208}
  \end{eqnarray}
  We need a first-order expansion of $S(x,\xt+\dxt)$:
  \begin{equation} \label{50}
      S(x,\xt+\dxt) = 
      S(x,\xt) + \frac{\xt-x+(g(\xt)-h(x))g'(\xt)}{S(x,\xt)}\dxt + o(\dxt).
  \end{equation}
  From Corollary \ref{cor1} and equation \eqref{14}, we get that
  \begin{equation} \label{51}
      \xt-x+(g(\xt)-h(x))g'(\xt) 
      =
      \left( -P_0 + g(\xt) - h(x) \right) g'(\xt)
      =
      -S(x,\xt) g'(\xt) .
  \end{equation}
  Substituting this into \eqref{50} and using Theorem \ref{thm2}, we get
  \begin{equation} \label{52}
      S(x,\xt+\dxt) = 
      S(x,\xt) - g'(\xt)S(x,\xt)\theta + o(\theta).
  \end{equation}
  Adding terms and using \eqref{14}, we get the claimed result.
\end{proof}

\begin{lemma} \;\;\; \label{thm5}
  $\xtt = \xt + \left(S(x,\xt) + g(\xt) A(\xt) \right) \theta + o(\theta) $.
\end{lemma}
\begin{proof}
  Simple geometry followed by application of previously derived first-order
  expansions reveal that
  \begin{eqnarray}
    \xtt & = & \xt + \dxt + g(\xt + \dxt) \tan \thetat  \nonumber \\[0.2in]
         & = & \xt + \dxt + \left( g(\xt) + g'(\xt) \dxt \right) \thetat +
		 o(\thetat) \nonumber \\[0.2in]
	 & = & \xt + \left(S(x,\xt) + g(\xt) A(\xt) \right) \theta + o(\theta) .
  \end{eqnarray}
  This completes the proof.
\end{proof}

Now consider a 
focusing element placed over the interval $[-\halfd,\halfd]$ of
the positive $x$-axis of focal length $f$ as shown in Figure 
\ref{fig:twomirrors}.
The electric field at the image plane is
\begin{equation} \label{101}
    E(\xi) 
    = \int_a^b e^{-ik\left( \frac{\xtt}{f}\xi - P(x)+P_0+y_0 \right)} dx .
\end{equation}
where $k = 2 \pi/\lambda$ is the wave number.  
Substituting the first order expansions for $\xtt$ and $P(x)$, we get
\begin{eqnarray}
    E(\xi) 
    & = &
    \int_{a}^{b} e^{
	         -ik \left(
		       \frac{\xt(x) 
		             + \left( S(x,\xt(x)) 
			            + g(\xt(x))A(\xt(x)) \right) 
			       \theta
			    }{f}
			    \xi
		       -
		       x \theta
		     \right)
    	       }
	       dx
    \nonumber \\
    & = &
    \int_{-\halfd}^{\halfd} e^{
	         -ik \left(
		       \frac{\xt 
		             + \left( S(x(\xt),\xt) 
			            + g(\xt)A(\xt) \right) 
			       \theta
			    }{f}
			    \xi
		       -
		       x(\xt) \theta
		     \right)
    	       }
	       A\left(\xt\right) d\xt . \label{102}
\end{eqnarray}
The second integral follows from the substitution $x = x(\xt)$.
%(Note: The physically astute reader might object that \eqref{101} and
%\eqref{102} seem to indicate that conservation of
%energy does not hold between the entrance and exit pupils.
%However, conservation of energy is a statement about two-dimensional 
%mirrors in three-dimensional space.  We plan to address this higher
%dimensional case in a future paper.  In that paper, we will check 
%conservation of energy between the two pupils.)

\subsection{Recasting in Unitless Terms}

To write the electric field using unitless quantities, we introduce
$\thetab$, $\thetatb$, $\xib$, and $\xb$ defined as follows:
%\begin{eqnarray}
%    \theta & = & \thetab \frac{\lambda}{D} 
%%                 = \thetab \frac{\lambda}{d \langle A \rangle}
%		 \\[0.2in]
%    \xi    & = & \xib \frac{f \lambda}{d} \\[0.2in]
%    \xt    & = & \xb d .
%\end{eqnarray}
\begin{equation}
    \theta = \thetab \frac{\lambda}{D} ,
    \qquad
    \tilde{\theta} = \thetatb \frac{\lambda}{d} ,
    \qquad
    \xi    = \xib \frac{f \lambda}{d} ,
    \qquad
    \xt    = \xb d 
\end{equation}
where $D=b-a$ denotes the aperture of the primary mirror.
Each of these quantities is unitless.  
%Angle $\thetab$ is measured in terms of the usual $\lambda/D$ measure.
Associated with these changes of units, we introduce the following
normalized versions of the apodization function, the transfer function, etc.:
\begin{eqnarray}
    \boldsymbol{A}(\xb) & = & A(\xb d) d / D \\
    \boldsymbol{x}(\xb) & = & x(\xb d)/D \\
    \boldsymbol{S}(\xb) & = & S(x(\xb d), \xb d) / D \\
    \boldsymbol{g}(\xb) & = & g(\xb d) / d .
\end{eqnarray}
Making these substitutions, we get
\begin{equation}
    E(\xib)
    = 
    D \int_{-\half}^{\half} e^{
	         -2 \pi i \left(
		       \xb \xib + \left( \boldsymbol{S}(\xb)
			            + \boldsymbol{g}(\xb) \boldsymbol{A}(\xb) \right) 
			       \frac{\lambda}{d}
			       \thetab \xib 
		       -
		       \boldsymbol{x}(\xb) \thetab 
		     \right)
    	       }
	       \boldsymbol{A}(\xb) d\xb
\end{equation}
The term in the exponential with the $\lambda/d$ factor is orders of magnitude
smaller than the other terms, hence we drop it.  The result is an explicit
expression for the electric field in the image plane as a function of
incidence angle $\thetab$, which is valid for small $\thetab$:
\begin{theorem} \;\;\; \label{thm6}
    To first order in $\thetab$,
    $
    E(\xib)
    = 
    \ds
    D \int_{-\half}^{\half} 
        e^{ -2 \pi i \left( \xb \xib 
	                   - \boldsymbol{x}(\xb) \thetab 
		     \right) } \boldsymbol{A}(\xb) d\xb .
    $
%    denotes a unitless representation of the transfer function
%    and
%    is the apodization function defined over a unit domain 
%    $\xb \in [-\half,\half]$ and having unit area.
\end{theorem}

It is easy to check that the normalized apodization function is related
to the normalized transfer function in the same way as before normalization:
\begin{equation}
    \frac{d\boldsymbol{x}}{d\xb}(\xb) = \boldsymbol{A}(\xb) .
\end{equation}
Also, the domain of the $\boldsymbol{A}(\cdot)$ is $[-\half,\half]$ and it 
integrates to one.

The magnitude of the complex electric field is easy to compute if we
assume that $\boldsymbol{A}(\cdot)$ is an even function.  With this assumption,
it follows that
\begin{equation} \label{30}
   \boldsymbol{x}(\xb) - \boldsymbol{x}(0) = \int_0^{\xb} \boldsymbol{A}(u) du
\end{equation}
is an odd function. % (and that $x(0) = (a+b)/2$).  
Hence,
\begin{equation} \label{31}
    E(\xib)
    = 
    D e^{ 2 \pi i \boldsymbol{x}(0) \thetab}
    \int_{-\half}^{\half} 
       \cos\left( 2 \pi 
          \left( \xb \xib - (\boldsymbol{x}(\xb)-\boldsymbol{x}(0)) \thetab \right) \right) 
       \boldsymbol{A}(\xb) d\xb .
\end{equation}

The point-spread function (psf), which is the square of the magnitude of the 
electric field, is then given to first order by
\begin{equation} \label{26}
    \Psf(\xib)
    = 
    D^2 
    \left(
    \int_{-\half}^{\half} 
       \cos\left( 2 \pi 
           \left( \xb \xib - (\boldsymbol{x}(\xb)-\boldsymbol{x}(0)) \thetab \right) \right) 
       \boldsymbol{A}(\xb) d\xb 
    \right)^2 .
\end{equation}

\subsection{Nonlinear Effects}

If the expression $\boldsymbol{x}(\xb)-\boldsymbol{x}(0)$ is linear, 
then the off-axis psf is just
a shift of the on-axis psf.  Deviation from linearity, on the other hand,
produces image degradation.  One measure of such deviation is the rms phase
error relative to $\boldsymbol{A}(0)$:
\begin{equation}
    \mbox{Phase-Error}
    =
    \left(
        \int_{-\half}^{\half}
	    \left(
	        \boldsymbol{x}(\xb) - \boldsymbol{x}(0) - \boldsymbol{A}(0)\xb
	    \right)^2
	d\xb
    \right)^{1/2} .
\end{equation}

It is easy to check that the unitless first-order relationship between
input angle $\thetab$ and output angle $\thetatb$ is 
\begin{equation}
    \thetatb = \boldsymbol{A}(\xb) \thetab .
\end{equation}
For a uniform apodization, $\boldsymbol{A}(\xb) = 1$ (see next section)
and hence the unitless input and output angles match.  
However, for a nonuniform apodization, the unitless output angle 
depends on position across the pupil.  On average, it equals the unitless
input angle.  But, the physical reality dictates using an amplitude weighted
average rather than a simple average.  In this case, the average unitless
output angle is larger than the input angle:
\begin{equation}
    \langle \thetatb \rangle
    =
    \int_{-\half}^{\half}
    \thetatb dx
    =
    \int_{-\half}^{\half}
    \thetatb \boldsymbol{A}(\xb) d\xb
    =
    \thetab 
    \int_{-\half}^{\half}
    \boldsymbol{A}(\xb)^2 d\xb
    \ge
    \thetab 
\end{equation}
(this last inequality follows from the Cauchy-Schwarz inequality and
the fact that $\boldsymbol{A}(\xb)$ integrates to one).   For the types
of apodizations arising in high-contrast imaging, the average
unitless output angle can be more than twice as large as the input angle
(see the numerical example in Section \ref{sec:numex}).  This extra
magnification is one of the attractive features of this approach to
apodization.

\section{Simple Examples}
\label{sec:simpex}

In this section, we consider a few simple examples.  
%We assume that $d=1$ so that $\xb$ and $\xt$ are the same.

\subsection{Uniform Apodization}

If the apodization function $A(\cdot)$ is a constant, call it $A$,
then $x(\xt) = x_0 + A \xt$ and its inverse is $\xt(x) = (x-x_0)/A$.
Plugging these into \eqref{13}, we get that
\begin{eqnarray}
    h'( x ) & = & \frac{(1-\frac{1}{A})x + \frac{1}{A} x_0}{P_0} \\
    g'(\xt) & = & \frac{x_0}{P_0} + \frac{A-1}{P_0} \xt 
\end{eqnarray}
and hence that each mirror is parabolic.  In the special case where $A=1$,
the two mirrors are actually planar and the system is a simple {\em periscope}.
For $A \ne 1$, the differential equations integrate as follows:
\begin{eqnarray}
    h( x ) & = & c + \frac{(x-\xi)^2}{4H} - \frac{P_0}{2} \\
    g(\xt) & = & c + \frac{(\xt-\xi)^2}{4G}
\end{eqnarray}
where $c$ is an arbitrary constant of integration,
$\xi = -x(0)/(A-1)$ is the $x$-coordinate of the centerline of the
system, $H = AP_0/(2(A-1))$ is the focal length of the ``$h$'' mirror,
and $G = P_0/(2(A-1))$ is the focal length of the ``$g$'' mirror.
Note that $H = A G$ and hence that the magnification of the system is $A$.
Using the fact that $A = D/d$,
the normalized apodization function is 
\begin{equation}
    \boldsymbol{A}(\xb) = A(\xb d) \frac{d}{D} = Ad/D = 1
\end{equation}
and the normalized transfer function is 
\begin{equation}
    \boldsymbol{x}(\xb) = \frac{x(\xb d)}{D} = \frac{x_0}{D} + \xb .
\end{equation}
Since $\boldsymbol{x}(\xb)$ depends linearly on $\xb$, 
it follows that the off-axis
point-spread function is just a shift of the on-axis psf:
\begin{equation} \label{60}
    \Psf(\xib)
    = 
    D^2 
    \left(
    \int_{-\half}^{\half} 
       \cos\left( 2 \pi \xb \left( \xib - \thetab \right) \right) 
       d\xb 
    \right)^2 .
\end{equation}
Of course, in practice such a pair of parabolic mirrors will
exhibit off-axis errors beyond
just a shift of the psf but these errors can only be captured by a 
second-order (or higher) analysis.

\subsection{A Cosine Apodization}

A simple explicit apodization function that approximates the sort of smoothly
tapering apodizations needed for high-contrast imaging is given by the cosine
function: 
\begin{equation}
    A(\xt) = a \left( 1 + \cos \frac{2 \pi \xt}{d} \right).  
\end{equation}
For this case, the transfer function
$x(\xt)$ is given by 
\begin{equation} \label{61}
    x(\xt) = x(0) + a \xt + a \frac{d}{2\pi} \sin \frac{2 \pi \xt}{d}
\end{equation}
and the ``$g$'' mirror's surface is given by
\begin{equation} \label{62}
    g(\xt) = c + \frac{x(0)\xt + \frac{a-1}{2} \xt^2 
               - a \frac{d^2}{4\pi^2} \cos(2 \pi \xt)}{P_0}.
\end{equation}
The expression for the ``$h$'' mirror's surface can be given explicitly
using \eqref{25} and the inverse of $x(\xt)$ but it is messy so we don't
record it here.
The amplitude parameter $a$ determines the relationship between $d$ and $D$:
\begin{equation}
    \frac{D}{2} = x(\halfd) - x(0) = a \frac{d}{2}
    \qquad \Longrightarrow \qquad
    D = a d .
\end{equation}
The normalized apodization is
\begin{equation}
    \boldsymbol{A}(\xb) = 1 + \cos(2 \pi \xb)
\end{equation}
and the normalized transfer function is
\begin{equation}
    \boldsymbol{x}(\xb) = 
    \frac{x_0}{D} + \xb + \frac{1}{2 \pi} \sin(2 \pi \xb) .
\end{equation}
The nonlinearity of the transfer function implies a nontrivial transformation
in addition to the usual shift of the off-axis psf:
\begin{equation} \label{63}
    \Psf(\xib)
    = 
    D^2 
    \left(
    \int_{-\half}^{\half} 
       \cos\left( 2 \pi \xb \left( \xib - \thetab \right) 
                  - \thetab \sin(2 \pi \xb) \right) 
       \boldsymbol{A}(\xb) d\xb 
    \right)^2 .
\end{equation}

In the next section, we consider a similar smooth apodization and plot
some of the off-axis psf's to see explicitly the impact of this nonlinear
effect.

\section{A Numerical Example}
\label{sec:numex}

\begin{figure}
\begin{center}
%\text{\includegraphics[angle=-90,origin=c,width=4.0in]{C:/user/rvdb/public_html/JAVA/tpf/apodmirror/mask.pdf}}
\text{\includegraphics[width=4.0in]{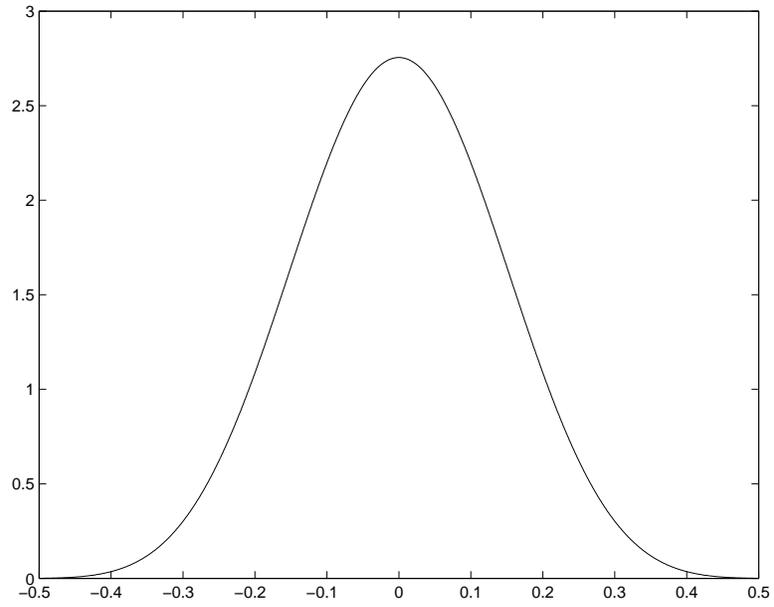}}
\end{center}
\caption{A unit area apodization providing contrast of $10^{-10}$ from $4
\lambda/D$ to $60 \lambda/D$.}
\label{fig:apod}
\end{figure}

\begin{figure}
\begin{center}
%\text{\includegraphics[width=4.0in]{C:/user/rvdb/public_html/JAVA/tpf/apodmirror/psf0.pdf}}
\text{\includegraphics[width=4.0in]{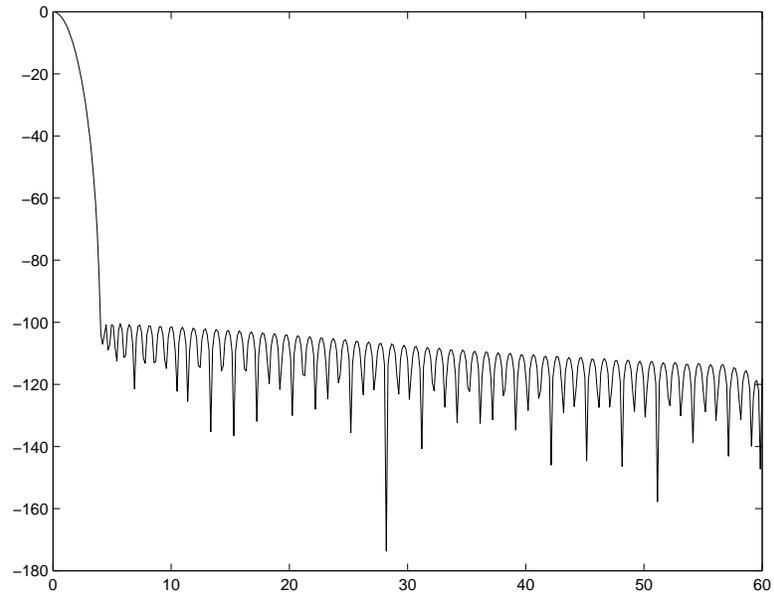}}
\end{center}
\caption{The on-axis point spread function for the apodization shown in Figure
\ref{fig:apod}.}
\label{fig:onaxis}
\end{figure}

The apodization function shown in Figure \ref{fig:apod} has unit area 
and provides $10^{-10}$ contrast from $4 \lambda/D$ to $60 \lambda/D$.
It was computed using the methods given in \cite{VSK03}.
The on-axis point spread function, computed by setting $\thetab = 0$ in
\eqref{26} is shown in Figure \ref{fig:onaxis}.

\begin{figure}
\begin{center}
%\text{\includegraphics[width=4.0in]{C:/user/rvdb/public_html/JAVA/tpf/apodmirror/psf0_02.pdf}}
\text{\includegraphics[width=4.0in]{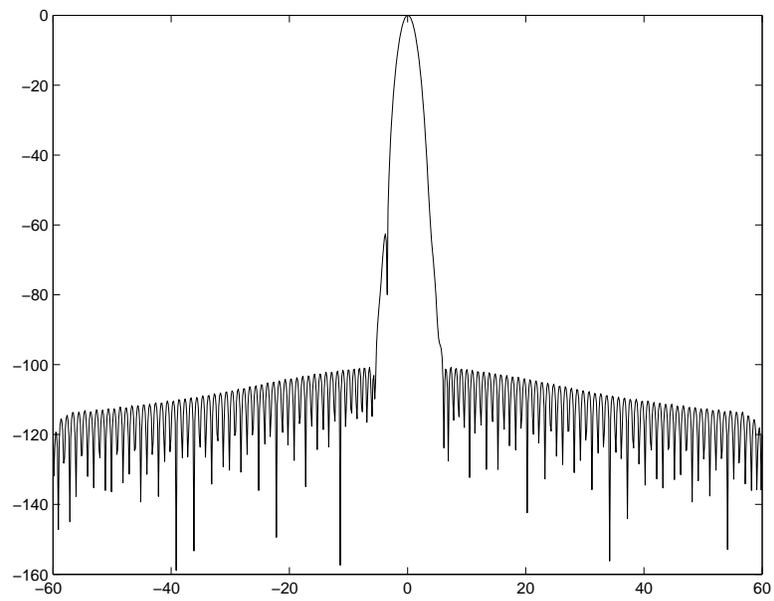}}
\end{center}
\caption{The off-axis point spread function for the apodization shown in Figure
\ref{fig:apod} computed at $\theta = 0.02 \lambda/D$.}
\label{fig:offaxis}
\end{figure}

In a solar-like system, a star's diameter is about 0.01au.  Hence,
if a $1$au Earth-like planet appears from Earth at $5 \lambda/D$, then
the angular extent of the star would be from 
$-0.025 \lambda/D$ to $0.025 \lambda/D$.  Therefore, starlight will enter
the system from angle $\thetab = 0.02 \lambda/D$.  The psf associated
with this slightly off-axis starlight is shown in Figure \ref{fig:offaxis}.
Although it is not possible to see it from the figure, the threshold for 
high contrast, i.e. $10^{-10}$, is pushed
to $6.0\lambda/D$.  Thresholds for high-contrast associated with
other values of $\thetab$ are shown in Table \ref{tbl1}.
It is clear from this table that non-zero stellar size and pointing 
error increase the inner working angle to at least $6 \lambda/D$.

\begin{figure}
\begin{center}
%\text{\includegraphics[width=4.0in]{C:/user/rvdb/public_html/JAVA/tpf/apodmirror/psf5.pdf}}
\text{\includegraphics[width=4.0in]{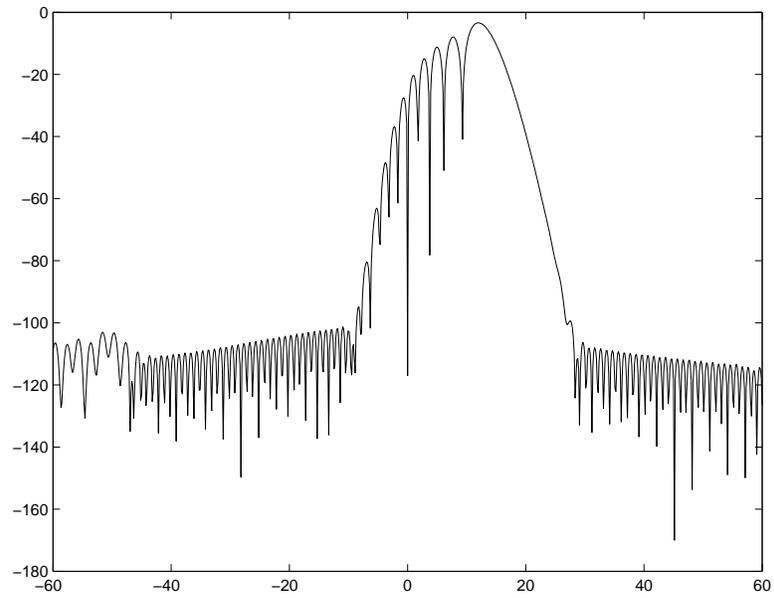}}
\end{center}
\caption{The off-axis point spread function for the apodization shown in Figure
\ref{fig:apod} computed at $\theta = 5\lambda/D$. Note that the main lobe
appears at $12 \lambda/D$.  The nonuniformity of the apodization accounts for
both the spreading out of the point spread function as well as the apparent
magnification by a factor of $12/5 = 2.4$.}
\label{fig:offaxis5}
\end{figure}

\begin{figure}
\begin{center}
%\text{\includegraphics[width=4.0in]{C:/user/rvdb/public_html/JAVA/tpf/apodmirror/psf10.pdf}}
\text{\includegraphics[width=4.0in]{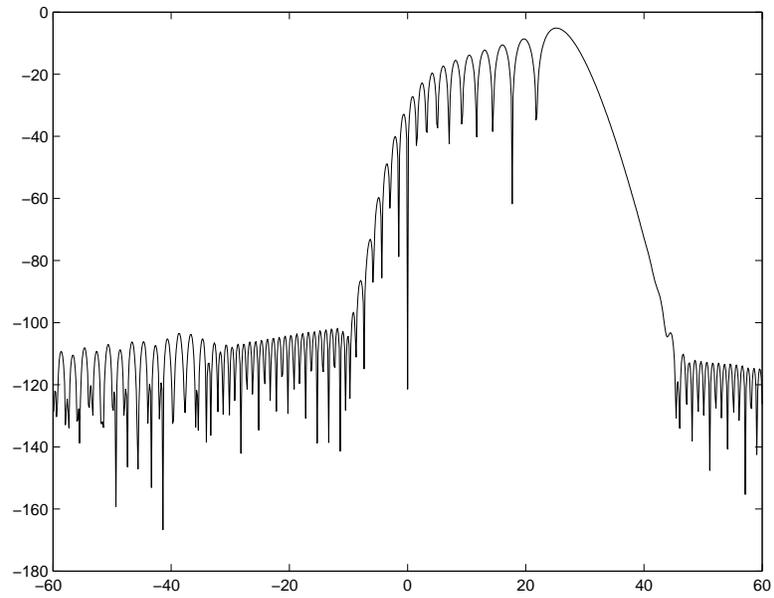}}
\end{center}
\caption{The off-axis point spread function for the apodization shown in Figure
\ref{fig:apod} computed at $\theta = 10\lambda/D$. Note that the main lobe
appears at about $22 \lambda/D$ for an effective magnification of about $2.2$.}
\label{fig:offaxis10}
\end{figure}

The psf's associated with a planet at $5$ and $10$ $\lambda/D$ are shown 
in Figures \ref{fig:offaxis5} and \ref{fig:offaxis10}.
Note that the magnification due to nonlinearity, described at the end of
Section \ref{sec:oap}, is clearly evident.  For example, in Figure
\ref{fig:offaxis5}, the input angle is $5 \lambda/D$ but the peak of the psf
occurs at about $12 \lambda/D$.  Hence, there is a magnification by more than
a factor of two.

On the other hand,
the unsharpening of the psf's and the corresponding reduction in peak
value suggest that a contrast level of $10^{-10}$ for the on-axis
case might not be sufficient for planet detection using only the two-mirror
system discussed here.  \citet{Guy03} suggests that an additional set of two
mirrors can be used to restore the off-axis (planet) images with good
fidelity.  The detailed investigation of this addition to the optics will be
investigated in a future paper.

%\acknowledgements
{\bf Acknowledgements.}
We would like to express our gratitude to N.J. Kasdin, D.N. Spergel, and E.
Turner for the many enjoyable and stimulating discussions in regard to this
work.  
In particular, N.J. Kasdin read several drafts carefully and provided
helpful advice as we struggled with various parts of this paper.
This research was partially performed for the 
Jet Propulsion Laboratory, California Institute of Technology, 
sponsored by the National Aeronautics and Space Administration as part of
the TPF architecture studies and also under contract number 1240729.
The second author received support from the NSF (CCR-0098040) and
the ONR (N00014-98-1-0036).

\bibliography{../lib/refs}   %>>>> bibliography data

\begin{thebibliography}{5}
\expandafter\ifx\csname natexlab\endcsname\relax\def\natexlab#1{#1}\fi
\expandafter\ifx\csname url\endcsname\relax
  \def\url#1{{\tt #1}}\fi

\bibitem[Goncharov et~al.(2002)Goncharov, Owner-Petersen, and Puryayev]{GOPP02}
A.~Goncharov, M.~Owner-Petersen, and D.~Puryayev.
\newblock Apodization effect in a compact two-mirror system with a spherical
  primary mirror.
\newblock {\em Opt. Eng.}, 41\penalty0 (12):\penalty0 3111, 2002.

\bibitem[Guyon(2003)]{Guy03}
O.~Guyon.
\newblock Phase-induced amplitude apodization of telescope pupils for
  extrasolar terrerstrial planet imaging.
\newblock {\em Astronomy and Astrophysics}, 404:\penalty0 379--387, 2003.

\bibitem[Labeyrie(1996)]{Lab96}
A.~Labeyrie.
\newblock Resolved imaging of extra-solar planets with future $10$-$100$km
  optical interferometric arrays.
\newblock {\em Astronomy and Astrophysics}, 118:\penalty0 517, 1996.

\bibitem[Traub(1986)]{Tra86}
W.A. Traub.
\newblock Combining beams from separated telescopes.
\newblock {\em Applied Optics}, 25:\penalty0 528--532, 1986.

\bibitem[Vanderbei et~al.(2003)Vanderbei, Spergel, and Kasdin]{VSK03}
R.J. Vanderbei, D.N. Spergel, and N.J. Kasdin.
\newblock 
%\href{http://www.princeton.edu/~rvdb/tex/starshape/ms.pdf}
{Circularly
  Symmetric Apodization via Starshaped Masks}.
\newblock {\em Astrophysical Journal}, 2003.
\newblock Submitted.

\end{thebibliography}
\bibliographystyle{plainnat}   %>>>> makes bibtex use natbib.bst

\clearpage

\clearpage

\begin{table}
\centering
\begin{tabular}{||r|rr|r|r||} \tableline
Sky & \multicolumn{4}{c||}{Focal Plane} \\ \tableline
$\theta$ & \multicolumn{2}{c|}{Contrast threshold} & angle of peak intensity &
	effective angular \\
$\lambda/D$ & \multicolumn{2}{c|}{$\lambda/D$} & 
	\multicolumn{1}{c|}{$\lambda/D$} & magnification
	$\lambda/D$ \\
%
%\emph{} & \emph{Contrast} \\
%\emph{$\theta$} & \emph{threshold} \\
%\emph{($\lambda/D$)} & \emph{($\lambda/D$)} \\

\tableline
0.00  & 4.0  & -4.0 & 0.0 & \\
0.01  & 5.2  & -5.2 & 0.023 & 2.3 \\
0.02  & 6.0  & -5.2 & 0.045 & 2.3 \\
0.10  & 6.4  & -6.0 &  0.23 & 2.3 \\
0.50  & 9.9  & -6.6 &  1.05 & 2.1 \\
2.00  & 15.9 & -7.7 &  4.35 & 2.2 \\
5.00  & 27.0 & -8.5 & 12.0  & 2.4 \\
10.00 & 43.5 & -9.3 & 25.2  & 2.5 \\
\tableline
\end{tabular}
%\caption{High-contrast ($10^{-10}$) threshold as a function of $\theta$.}
\caption{Output image properties as a function of input angle.
Column 1 is the input angle on the sky of a point source.
Columns 2 and 3 are the output angles at which the intensity in the focal
plane falls to $10^{-10}$ compared to the on-axis image peak.
Column 4 is the output angle at which the star image has its peak intensity.
Column 5 is the effective angular magnification factor, i.e., the ratio of
column 4 to column 1.}
\label{tbl1}
\end{table}

\end{document}